\documentclass[conference]{IEEEtran}

\usepackage{graphicx}
\usepackage{epstopdf} %
\usepackage{amssymb}
\usepackage{amsmath}
\usepackage{multirow}
\usepackage{comment}
\usepackage{algorithm}
\usepackage[]{algpseudocode}
\usepackage{balance}
\usepackage{mathtools}
\usepackage{url}  %
\usepackage{bm}
\usepackage[numbers,sort&compress]{natbib} %
\usepackage{xspace} %

\usepackage[labelformat=simple]{subcaption}

\makeatletter
\let\MYcaption\@makecaption
\makeatother
\usepackage[font=footnotesize]{subcaption}
\makeatletter
\let\@makecaption\MYcaption
\makeatother

\allowdisplaybreaks

\begin{document}
\title{Argus:~Smartphone-enabled Human Cooperation via Multi-Agent Reinforcement Learning for Disaster Situational Awareness}

\author{{Vidyasagar Sadhu, Gabriel Salles-Loustau, Dario Pompili, Saman Zonouz, Vincent Sritapan$^*$}\\
Department of Electrical and Computer Engineering, $^*$Cyber Security Division\\
Rutgers University, $^*$Department of Homeland Security Science \& Technology Directorate\\
\textit{ \{vidyasagar.sadhu, gs643, pompili, saman.zonouz\}@rutgers.edu}, \textit{vincent.sritapan@hq.dhs.gov}
}
\date{}
\maketitle

\thispagestyle{empty}
\pagestyle{plain}

\begin{abstract}
Argus exploits a Multi-Agent Reinforcement Learning~(MARL) framework to create a 3D mapping of the disaster scene using agents present around the incident zone to facilitate the rescue operations. The agents can be both human bystanders at the disaster scene as well as drones or robots that can assist the humans. The agents are involved in capturing the images of the scene using their smartphones (or on-board cameras in case of drones) as directed by the MARL algorithm. These images are used to build real time a 3D map of the disaster scene. Via both simulations and real experiments, an evaluation of the framework in terms of effectiveness in tracking random dynamicity of the environment is presented.

\end{abstract}

\section{Introduction}
Disaster management involves a very large number of heterogeneous agents in an uncontrolled and potentially hostile environment. These agents include victims and the rescue personnel that eventually intervene. Situational-awareness techniques involving better incident management strategies will help a long way reduce disaster damage. Several works focus on modeling incident response operations but very few can be used in real disasters as there are several challenges to be tackled in practical situations. Several challenges in designing an interactive incident response system such as the coordination of the intervention forces are presented in~\cite{Kyng:2006:CDI:1142405.1142450}. 

\textbf{Motivations: }Traditionally, incident reporting relies on expensive information collection from authorities. These reportings happen most often in an offline manner and fail to provide ahead of time actionable information to incident response team. However, in these situations it is common to find people present at those locations (bystanders/onlookers) to record the scene using their smartphones. Very few of the existing incident response frameworks take advantage of this valuable information. We argue that this information is valuable because it is first hand, real time and local, and could provide fine-grained information about the scene. Also, in some cases, this information can be complemented by drones taking pictures or videos of the disaster scene.

\textbf{Our Approach: }Contrary to existing solutions such as those based on CCTV cameras, our solution, \textit{Argus} (from \textit{Argus Panoptes}, the ``all-seeing" 100-eye giant in the Greek mythology), utilizes this valuable information coming directly from the bystanders and also provides a base framework wherein both the drones/robots and humans can collaboratively work to provide valuable and fine-grained information to the rescue authorities. Argus is implemented as a mobile application that can be installed on the users' smartphones to capture and process data such as images of the incident scene. We then consolidate this valuable data into an easy-to-visualize form such as a 3D map of the disaster scene in real time. The advantage of this approach over looking at individual images is that it is difficult to process the latter one by one as there might be a huge number of them streaming in. %
However, the same reasons become a boon for 3D reconstruction as it needs many images from different views. Also, generating a 3D model from a sequence of images is much cheaper than using other techniques such as 3D scanners. We believe that integrating other well-known technologies such as audio source separation into our framework can add value (such as to prioritize some regions over others).
In this work, we make use of a Multi-Agent Reinforcement Learning~(MARL) framework for adaptive and effective data collection. The reasons for considering MARL are as follows: (1) to enable full coverage of the scene, without which most of the images would correspond to a few conspicuous regions only (e.g., fire regions); (2) to enable fine-grained (high resolution in 3D maps) coverage in regions of interest specific to the user (e.g., behind the building).
 \begin{figure*}[ht!]
        \centering   
           \begin{subfigure}[t]{0.4\textwidth}
        		\centering
        		\includegraphics[width=.95\textwidth]{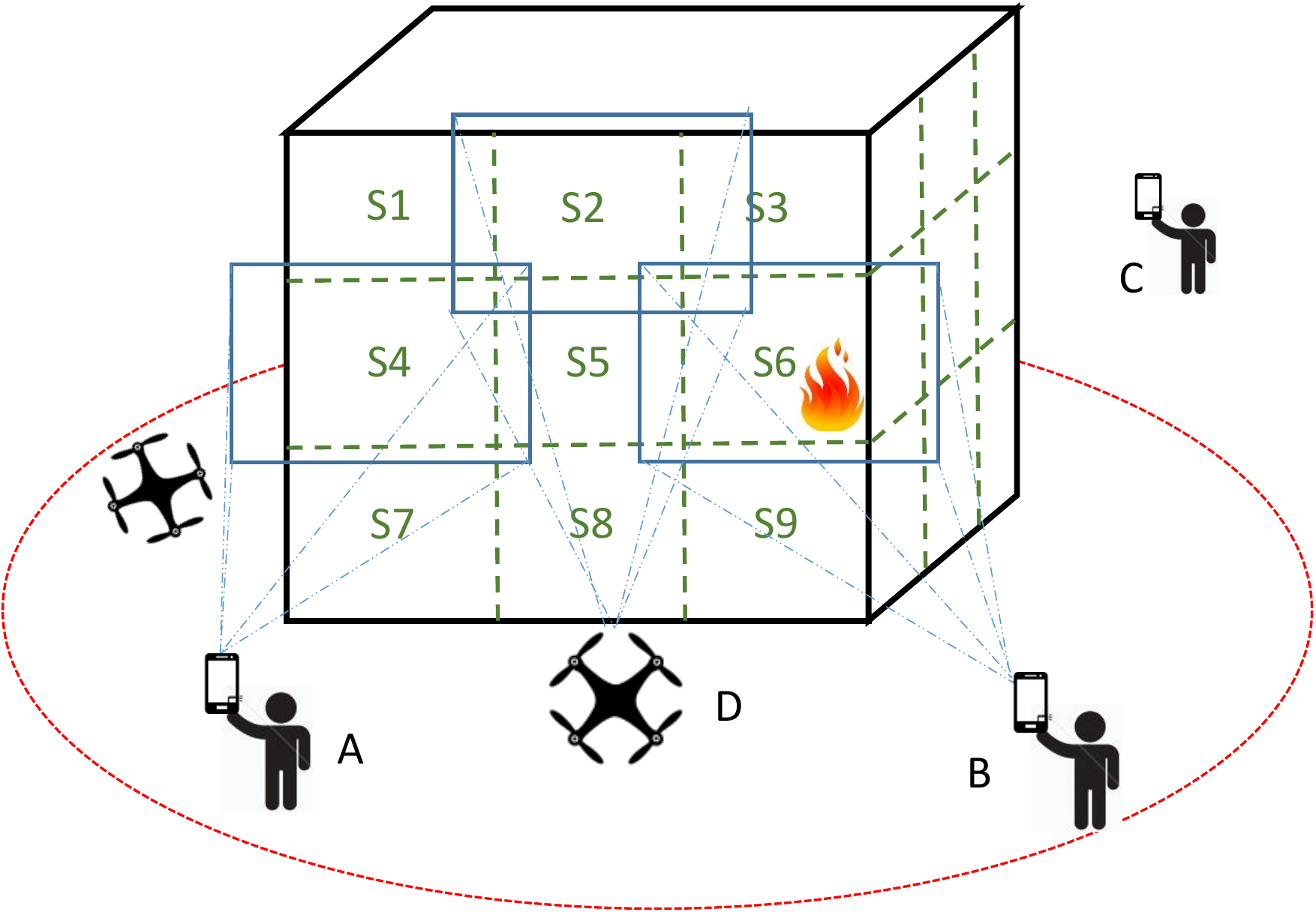}
        		\caption{Modelization of an incident zone with human agents~(A,B,C) and flying drones~(D). %
                }
        		\label{fig:argususecase}
        	\end{subfigure}%
            ~
        \begin{subfigure}[t]{0.6\textwidth}
        	\raisebox{10mm}
            \centering 
            \includegraphics[width=.95\textwidth]{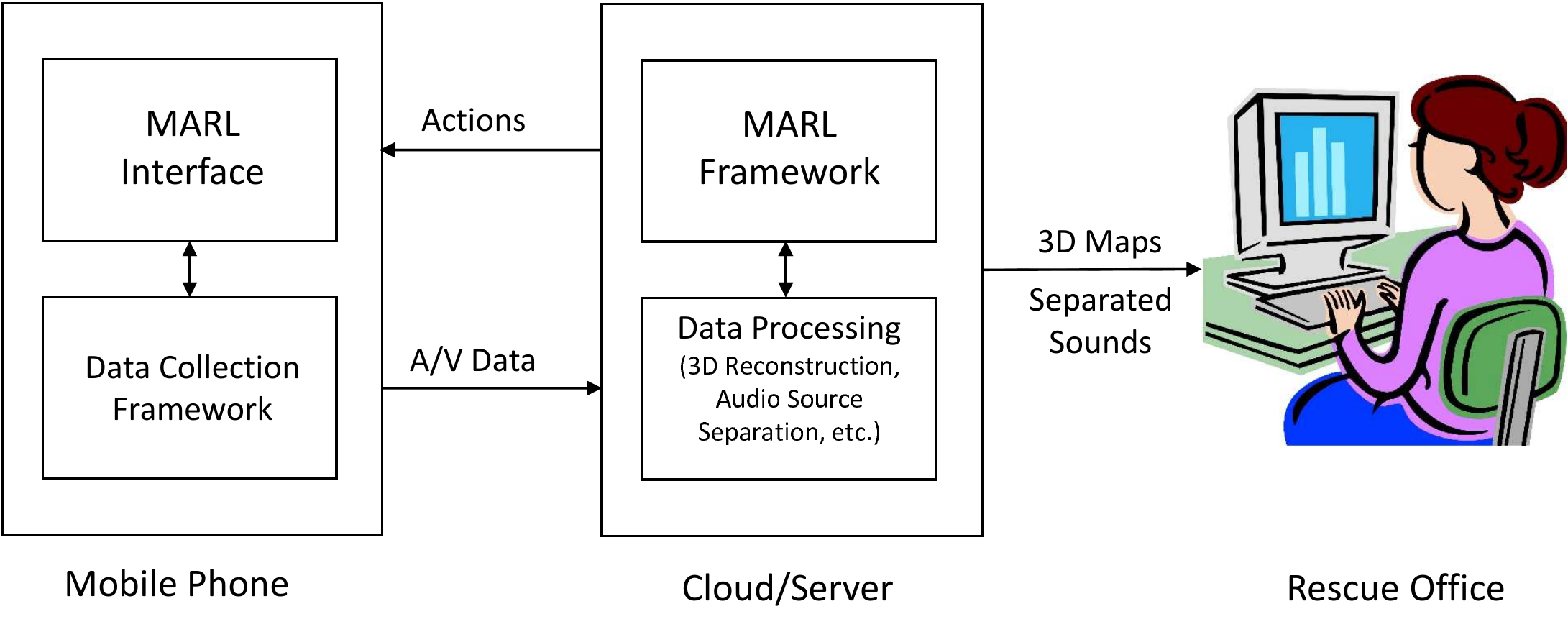}
            \caption{Argus High-level Architecture.}
            \label{fig:argusarch}
        \end{subfigure}
        \caption{\label{fig:argus} (a) Argus Use Case: a building on fire; (b) Argus High-level Architecture.}
        \vspace{-0.15in}
\end{figure*}
\\\indent\textbf{Related Work:}
The applicability and usefulness of MARL to build evacuation simulations for emergency/disaster situations and to model a large building containing multi-agent heterogeneous population attempting to evacuate in the presence of a non-stationary fire is investigated in~\cite{marlevac}.
In~\cite{ah2013}, design and development of a mobile system is presented that reconstructs 3D model of a building's interior structure in real time and fuses the visualization with the image of a thermal camera. RescueNet~\cite{rescuenet} proposes a networking paradigm based on MARL to enable reliable and high data-rate wireless multimedia communication among public safety agencies using mission policies that enable graceful degradation in the Quality of Service~(QoS) of the incumbent networks in times of emergency.
Various solutions for real-time monitoring of incident zones including interactive systems are proposed in~\cite{mehrotra2003project,Kyng:2006:CDI:1142405.1142450}. Design and implementation of a framework to simulate incident response such as building evacuation with modeling victims as agents with specific set of properties such as their visibility is presented in~\cite{simulation}. A multiple-agent Markov Decision Process~(MDP) model is leveraged to synchronize the actions of the response team after an incident in~\cite{ramchurn2015human}. None of these frameworks use MARL for data collection from the bystanders to create 3D maps as we do.
Regarding MARL, there are two types of frameworks, cooperative and competitive~\cite{marlsurvey}. In the former the agents work cooperatively towards a common goal (with same rewards), while in latter agents compete against each other with opposite rewards (e.g., the sum of the rewards of all agents is zero). Our scenario corresponds to the former one. A simple MARL framework for cooperative agents without assuming any coordination among the agents is proposed in~\cite{distqlearn1}. Each agent maintains a local Q-table and a policy, both of which are updated only if there is an improvement in the Q-values. However, for each local (state, action) pair, it requires that the other agents' actions happen infinitely often, which may not be practical in time-constrained scenarios. The idea of Q-value sharing is proposed in~\cite{distqlearn2,distqlearn3}. We make use of similar concept in our algorithm except that all agents share the same Q-table rather than having individual tables with knowledge sharing as they propose. 

\textbf{Our contributions} are summarized below:
\begin{itemize}
\item A model for real-time incident response data collection using Multi-Agent Reinforcement Learning~(MARL) (available as a mobile application) using humans as agents for data collection. 
\item Evaluation of the MARL framework through simulations to find optimal design parameters and to study its behavior for random fire occurrences. 
\end{itemize}

\textbf{Paper Outline: }In Sect.~\ref{sec:prop-soln}, we present the Argus architecture and the system design including the MARL and Q-learning frameworks. In Sect.~\ref{sec:evaluations}, we study our framework by varying the design parameters and also evaluate its performance for random fire occurrences. Finally, we conclude the paper and mention our future directions in Sect.~\ref{sec:discussions}.

\section{Proposed solution}\label{sec:prop-soln}

\begin{figure*}[t!]
        \centering   
           \begin{subfigure}[b]{0.23\textwidth}
        		\centering
        		\includegraphics[width=1\textwidth]{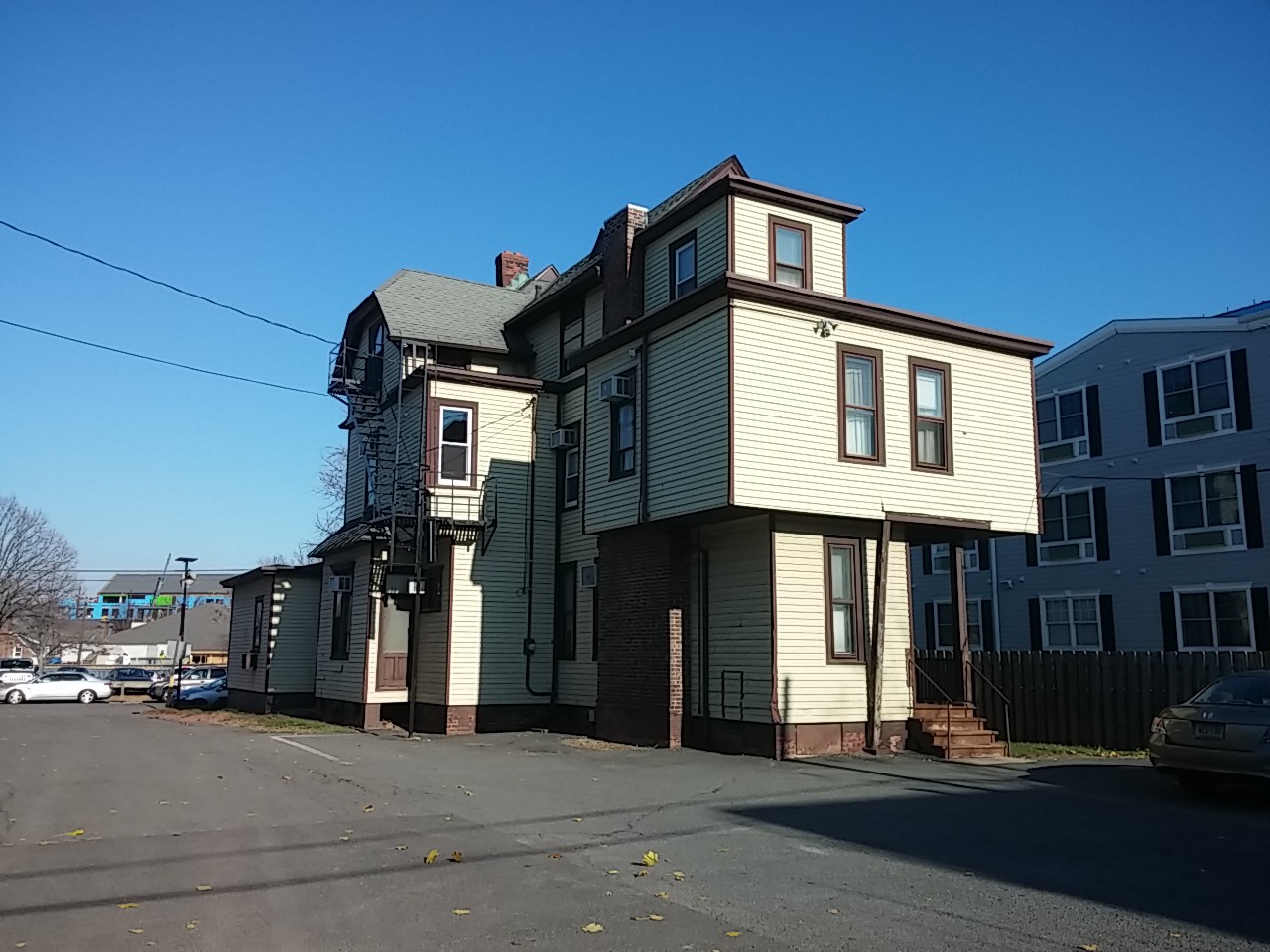}
        		\caption{}
        		\label{fig:orig}
        	\end{subfigure}
~
        \begin{subfigure}[b]{0.37\textwidth}  
            \centering 
            \includegraphics[width=1\textwidth]{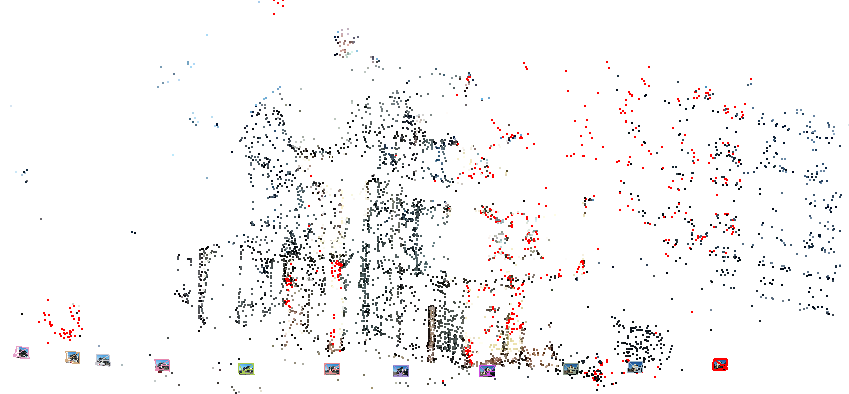}
            \caption{}
            \label{fig:sparse3d}
        \end{subfigure}
~
        \begin{subfigure}[b]{0.36\textwidth}   
            \centering 
            \includegraphics[width=1\textwidth]{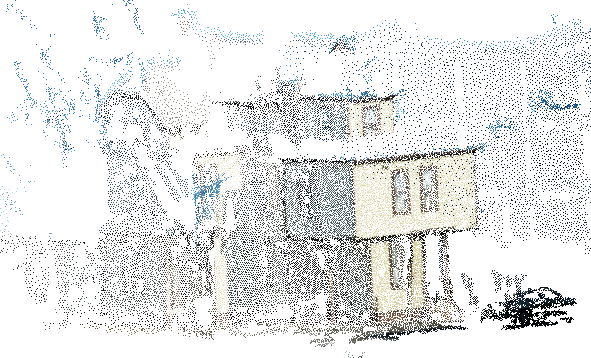}
            \caption{}
            \label{fig:dense3d}
        \end{subfigure}
        \caption{\label{fig:3dmap} (a) Original Image; (b) Sparse 3D reconstruction showing camera positions at the bottom; (c) Dense 3D reconstruction.}
        \vspace{-0.15in}
\end{figure*}

\textbf{Argus use case: }Fig.~\ref{fig:argususecase} presents a use-case scenario for Argus, which shows a building (represented by cube) on fire. Other use cases could be large forest fires, flooding, a terrorist attack, etc. Argus is installed as a mobile phone application and directs bystanders present at the incident zone on what actions to take (e.g., what pictures to take using their smartphones) to gather relevant information pertaining to the incident zone. This is achieved as follows. Argus decomposes the incident zone into multiple subzones with sufficient overlap. For example, in the case of building on fire, each subzone could correspond to a face or corner of the building with corners enabling overlap between the faces of the building. Each subzone is further divided into a rectangular grid consisting of states $S_0,...S_9$, as shown in Fig.~\ref{fig:argususecase}. Each grid is cast as a Markov Decision Process~(MDP), which is solved using MARL with agents being the bystanders around the incident zone (A,B,C) and any drones (D) that can assist humans in the data-gathering process. Argus mobile application, which interfaces with the MARL framework running in the cloud, directs the agents (humans or drones) on what information to collect from the incident zone by giving directions such as ``Left," ``Up," etc. The drones could take part in the process just as humans or could complement them by capturing information that may not be accessible to humans (for example, a forest lining one face of the building that is not accessible to humans) or they could be assigned special tasks to capture information in regions where it is incomplete and immediately needed.
As mentioned earlier, we have chosen to create a 3D map of the disaster scene, in which case the information gathered from the agents will be images or videos of the disaster scene. We aim at providing near real-time 3D mapping of the disaster scene so that the rescue personnel are aware of the entire situation, know where to immediately focus their attention on or plan appropriately. We used a GUI program called VisualsFM~\cite{visualsfm} for the 3D reconstruction, which consists of two parts, a well-known Structure From Motion~(SFM) algorithm~\cite{3dtut} for sparse 3D and CMVS/PMVS~\cite{3dtut} for dense 3D reconstruction. Fig.~\ref{fig:sparse3d} and Fig.~\ref{fig:dense3d}, respectively, show the sparse and the dense 3D reconstruction of a building (Fig.~\ref{fig:orig}) using SFM and PMVS/CMVS packages found in VisualSfM. We have used 12 images of the building taken from different views using LG~D800 smartphone with $1~\mathrm{MP}$ resolution. We would like to mention that in real incidents it is realistic assumption to have hundreds of images, making the reconstruction more detailed. Also, there will be a higher resolution in areas of interest such as fire regions, damage regions, etc. as they will be the reward states in MARL, as we explain below.

\textbf{Argus architecture: }Our framework consists of the following parts, as shown in Fig.~\ref{fig:argusarch}: 1. A client-side data-collection framework that is part of the mobile application for collecting audio and video from the users smartphone. 2. A server side, MARL framework that interfaces with the mobile application (MARL Interface) and directs the agents on what parts of the disaster scene to capture by providing directions such as ``Left," ``Up," etc. 
3. A server-side data-processing framework that uses the images and/or audio data sent by the clients to generate 3D maps and other processing tasks (such as audio source separation). This processed data in the form of 3D maps is then sent to rescue office personnel who are in a position to guide rescue personnel using this information. The MARL framework directs the agents (humans/drones) to capture photos in a proper way, i.e., capturing a few photos to give the context and many others primarily focused around the regions of interest. Our framework achieves this by assigning more rewards to the data corresponding to interest regions.
We make use of distributed Q-learning to solve our MARL problem. During the \emph{exploration phase} of Q-learning, the agents explore the entire disaster scene to find reward states and create a coarse 3D map of the entire scene. Then, during the \emph{exploitation phase}, more time is spent in reward states to create a finer 3D maps of those regions. We chose Q-learning as it is model free and can adapt to changing environments (such as fire propagation).

\textbf{System Design: }%
We now describe our MARL framework consisting of a distributed Q-learning approach for exploration and of a periodic exploration-exploitation mechanism to capture the dynamicity of the environment. To reiterate, our MARL framework is used for data collection only. As the images are obtained, 3D reconstruction is done using those images in real time, as mentioned in Sect.~\ref{sec:prop-soln}.

\underline{MARL Framework}: Although our framework is quite general and can be applied to any scenario, in order to be concrete we will continue with our example of building on fire as in Fig.~\ref{fig:argususecase}. We formulate the problem as a MDP as follows. 
We divide the building into 8 surface views consisting of 4 faces and 4 corners assuming a regular cuboid structure. Each surface and corner are modeled as a rectangular grid, and the bystanders present near that view are assigned to that particular MDP making that process a Multi-Agent System~(MAS). In case of corners, the grid is created by unfolding them to plain surfaces and considering some portions from each of the constituting faces. Each MDP acts independently of others. Images from all MDPs are fused to form the 3D map of the building. The \textit{states} are the cells (x,y) of the grid. For now, let us assume that the humans (agents) participating in our framework are static and can only pan or tilt their phones. An agent is said to be in a state (x,y) if it positions its smartphone camera focusing/centering on the cell (x,y). 
The \textit{actions} for each state are that of a standard grid-world MDP: \{Left, Right, Up, Down\}. Here, Left and Right are realized by \textit{panning} (rotating along vertical axis) the phone by a specific angle $\theta_x$ in those directions. Two continuous Right actions would mean panning by $2\theta_x$ to the right and so on. On similar lines, Up and Down are realized by \textit{tilting} (rotating along horizontal axis) the phone by a specific angle $\theta_y$. We could make use of gyroscopes in the phone to provide feedback to the users on when to stop panning (or tilting) the phone when they have covered an angle of $\theta_x$ (or $\theta_y$) in the specified directions. Whenever an agent takes an action to move to a different state, it captures the image of the current state (grid cell in the MDP) before panning/tilting the phone. This image determines the \textit{reward} the agent receives for that state. In other words, our reward is dependent on the state alone rather than on the state-action-state sequence. Since the regions with more fire are of primary interest, more rewards are assigned if more percentage of fire pixels is detected in the image. In order to avoid cheating, we normalize this by the zoom factor (Eq.~\ref{eq:reward}). We note that reward can be changed as per user's requirements. If uniform coverage is the final goal, incomplete regions will be reward states. Using this approach, the algorithm can gear the agents to capture images of interest to the user,
where our reward is,
\begin{equation}\label{eq:reward}
r(s) = \frac{\%\textsf{(fire-pixels)}}{\textsf{zoom-factor}}.
\end{equation}
Finally, the size of the grid and $\theta_x, \theta_y$ will be determined by these factors: the dimension of the incident zone, the average distance of the agents from the incident zone, and the average capture angle of the agents' devices.

\textit{1) Distributed Q-Learning: }In Q-learning the agent learns from each experience as it interacts with the environment and updates its Q-values based on the reward that it gets from this experience. Let us denote the set of all states $s$ as $S$, the set of all actions $a$ as $A$, and the reward as $r$. A single experience with the environment at time $t$ can be represented by the tuple $(s_t, a_t, r_t, s_{t+1})$. This means the agent was in state $s_t$, took action $a_t$, received reward $r_t$, and landed in state $s_{t+1}$. The agent learns (updates its Q-values) through a series of such experiences.
Let us denote the Q-value of a state-action pair at time $t$ as $Q_t(s_t,a_t)$. With the experience from $(s_t, a_t, r_t, s_{t+1})$, the Q-value can be updated as follows,
\begin{equation}\label{eq:Q}
Q_{t+1}(s_t,a_t) \leftarrow Q_{t}(s_t,a_t) + \alpha [r_t+\gamma max_{a}Q_{t}(s_{t+1},a) - Q_{t}(s_t,a_t)],
\end{equation}where $\alpha$ is the learning rate and $\gamma$ is the discount factor. As can be seen from~\eqref{eq:Q}, Q-values are updated based on the immediate reward, $r_t$, and on the optimal expected return, $\gamma max_{a}Q_{t}(s_{t+1},a)$. Note that the learning rate $\alpha$ is set to a large number in the beginning and is slowly reduced to ensure convergence.

Given that we have a MARL problem, we make use of distributed Q-learning (slightly modified from~\cite{distqlearn2,distqlearn3}) where the agents share the same Q-values using a shared database (with synchronous read/write) to expedite the exploration process. We would like to clarify that it is still a multi-agent system except that Q-values are same for all. Each agent still acts independently. As all agents are exploring in parallel and making use of the knowledge/experience gained from all other agents (Q-values), the exploration process is greatly sped up. 

\textit{2) Periodic Exploration: }There are two well-known exploration strategies: $\epsilon$-greedy approach and Boltzmann exploration. Even though both strategies drive the randomness in selecting action to reduce over time, there is a disadvantage with the former as $\epsilon$-greedy gives equal importance to all actions in case it needs to select a random action (with probability $\epsilon$) with no regard to the Q-values. Let us say there are 3 actions with decreasing Q-values other than the optimal action, which has even higher Q-value. In case of $\epsilon$-greedy, all 3 actions are given equal importance whereas it would be desirable to assign a probability that is directly proportional to the Q-value. Boltzmann exploration achieves this, where the probability of selecting an action is directly related to its Q-value (proportional to $e^{Q_t(s,a)/T}$),
\begin{equation}
\pi_t(s,a) = \frac{e^{Q_t(s,a)/T}}{\sum_{a'\in A}e^{Q_t(s,a')/T}}.
\end{equation}
Here, the temperature parameter, $T$, decides the amount of exploration or exploitation; if $T$ is large, all actions have almost equal probability resulting in pure exploration whereas when $T$ tends to 0 the optimal action has the highest probability, resulting in pure exploitation. 
\begin{figure}
  \centering
  \includegraphics[width=0.5\textwidth]{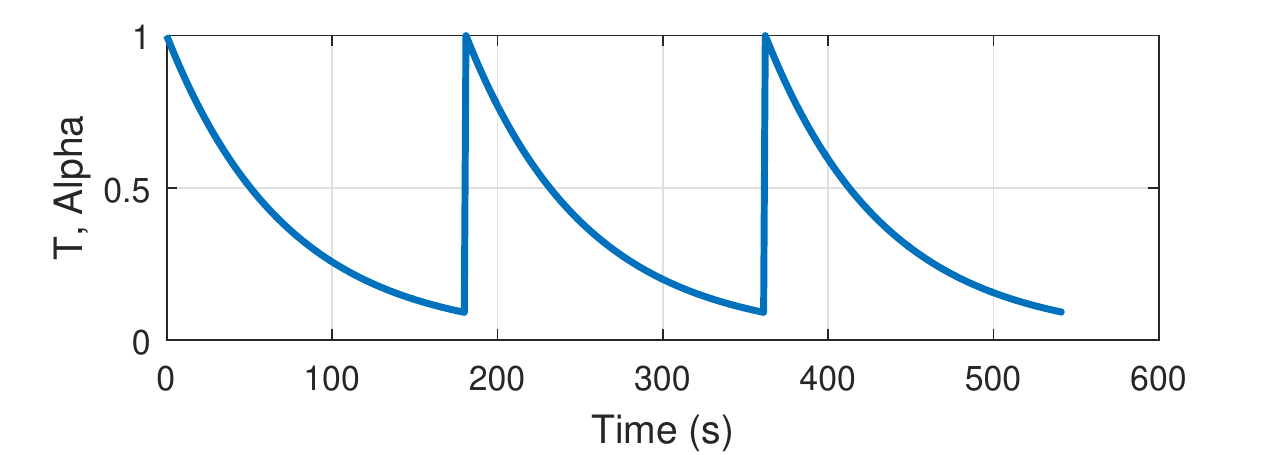}
  \caption{Variation of $T$, $\alpha$ to perform periodic exploration.}
  \label{fig:periodicexplrn}
  \vspace{-0.15in}
\end{figure}
In order to adapt to the dynamicity of the environment, we enable periodic exploitation. 
In each period, there is a short exploration phase followed by a long exploitation phase. Accordingly, we vary the exploration parameter ($T$) and the learning rate ($\alpha$) in a periodic fashion (Fig.~\ref{fig:periodicexplrn}). We note there are two extremes to this mechanism. On one extreme there is a need to account for random occurrences of fire in scenarios where the environment changes randomly, e.g., during a terrorist attack; in these cases, we reset the Q-values and the exploration parameter (Boltzmann temperature) across periods to enable new learning. On the other extreme, we have situations such as slow fire propagation where there is no significant change in the environment across periods; in this case, it is preferable to carry forward the Q-values to the next period so to leverage the learning in the previous period. The exploration parameter, $T$, too can be reduced quickly in the next period to enable less exploration and more exploitation. In both cases, the period will depend on the rate of dynamicity of the scene.
\begin{figure}[tp]
        \centering   
           \begin{subfigure}[t]{0.15\textwidth}
        		\centering
        		\includegraphics[width=1\textwidth]{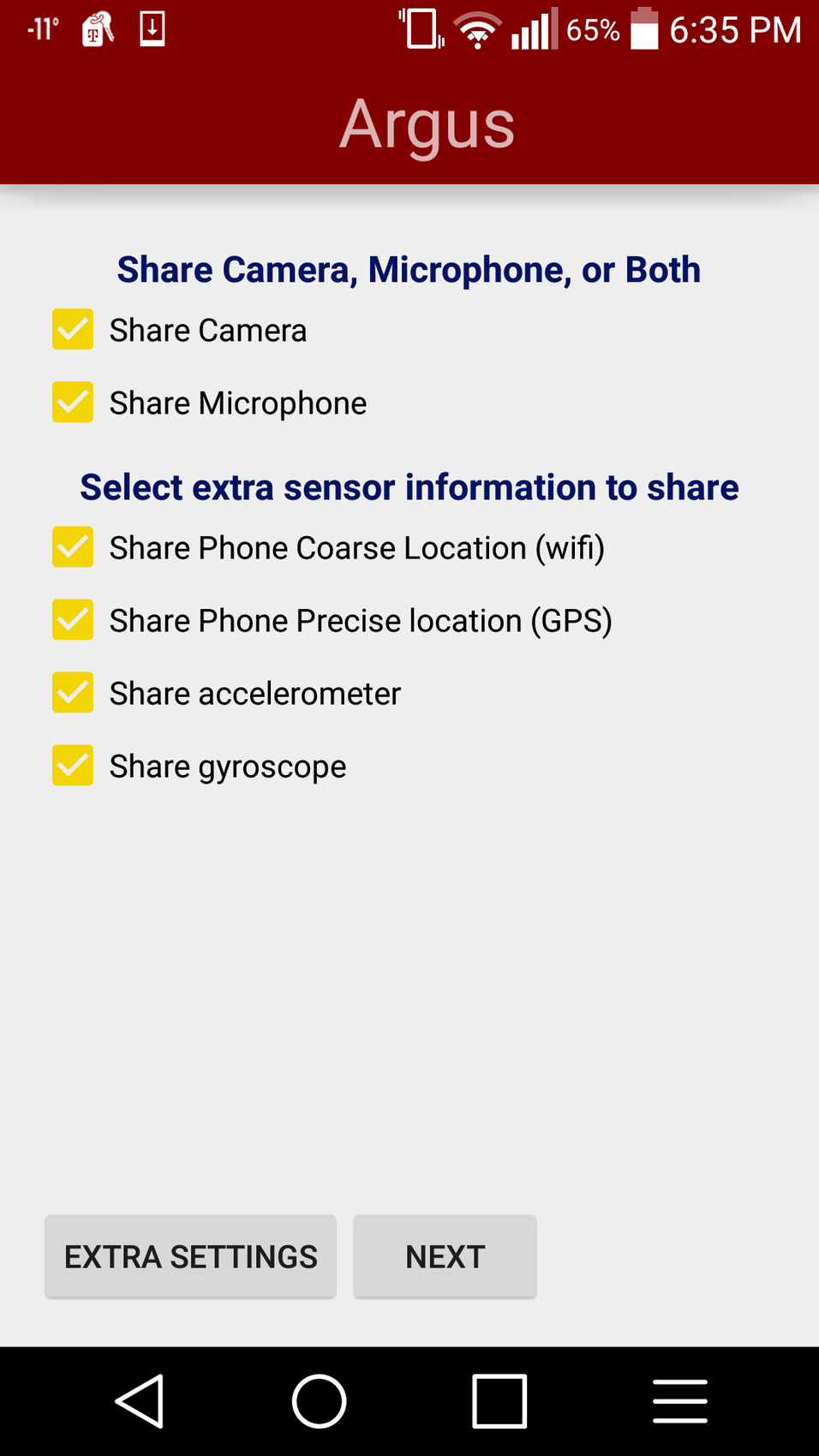}
        		\caption{}
        		\label{fig:argusconf}
        	\end{subfigure}
~
         \begin{subfigure}[t]{0.15\textwidth}   
             \centering 
             \includegraphics[width=\textwidth]{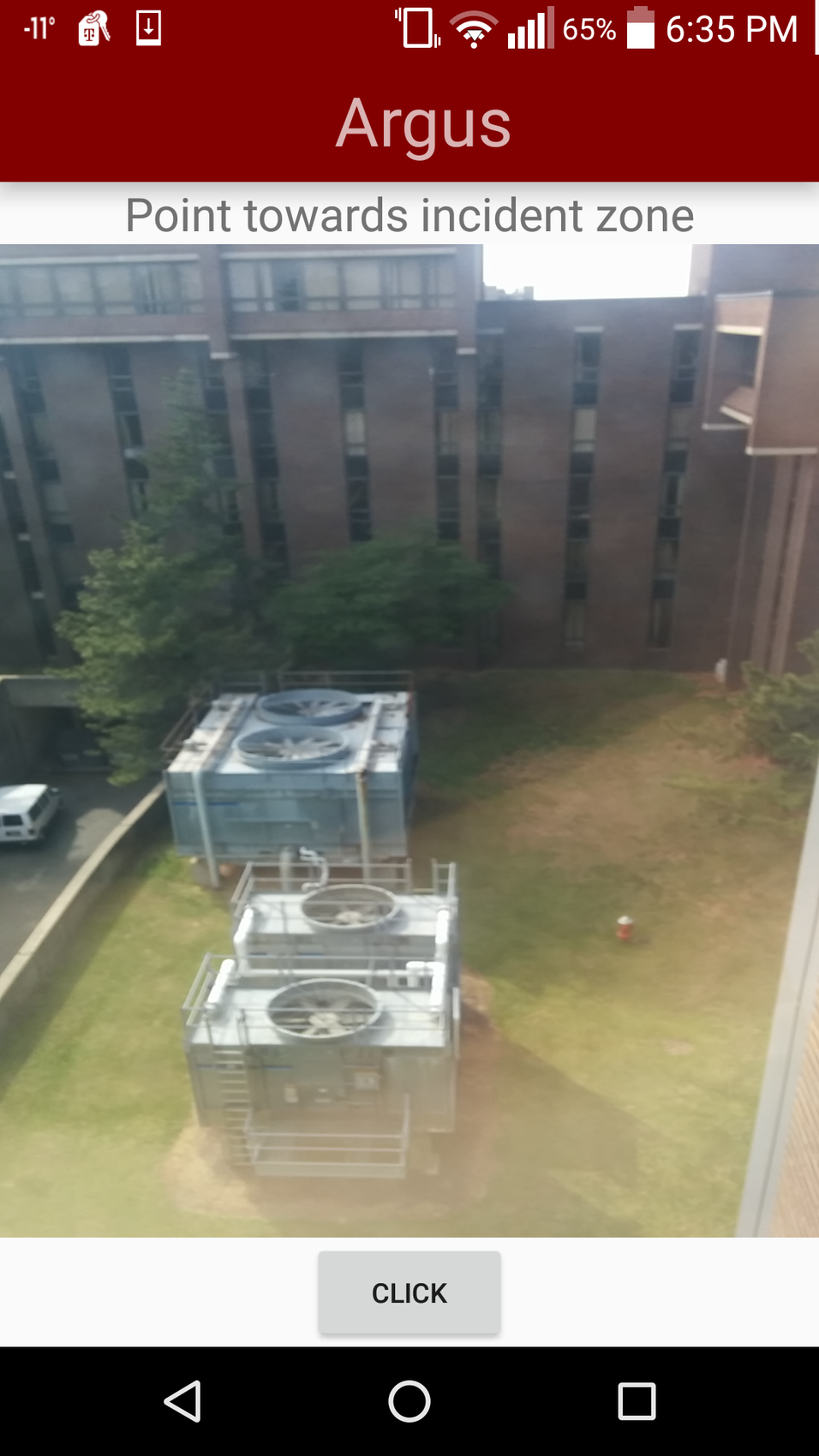}
             \caption{}
             \label{fig:arguspoint}
         \end{subfigure}
~
        \begin{subfigure}[t]{0.15\textwidth}  
            \centering 
            \includegraphics[width=1\textwidth]{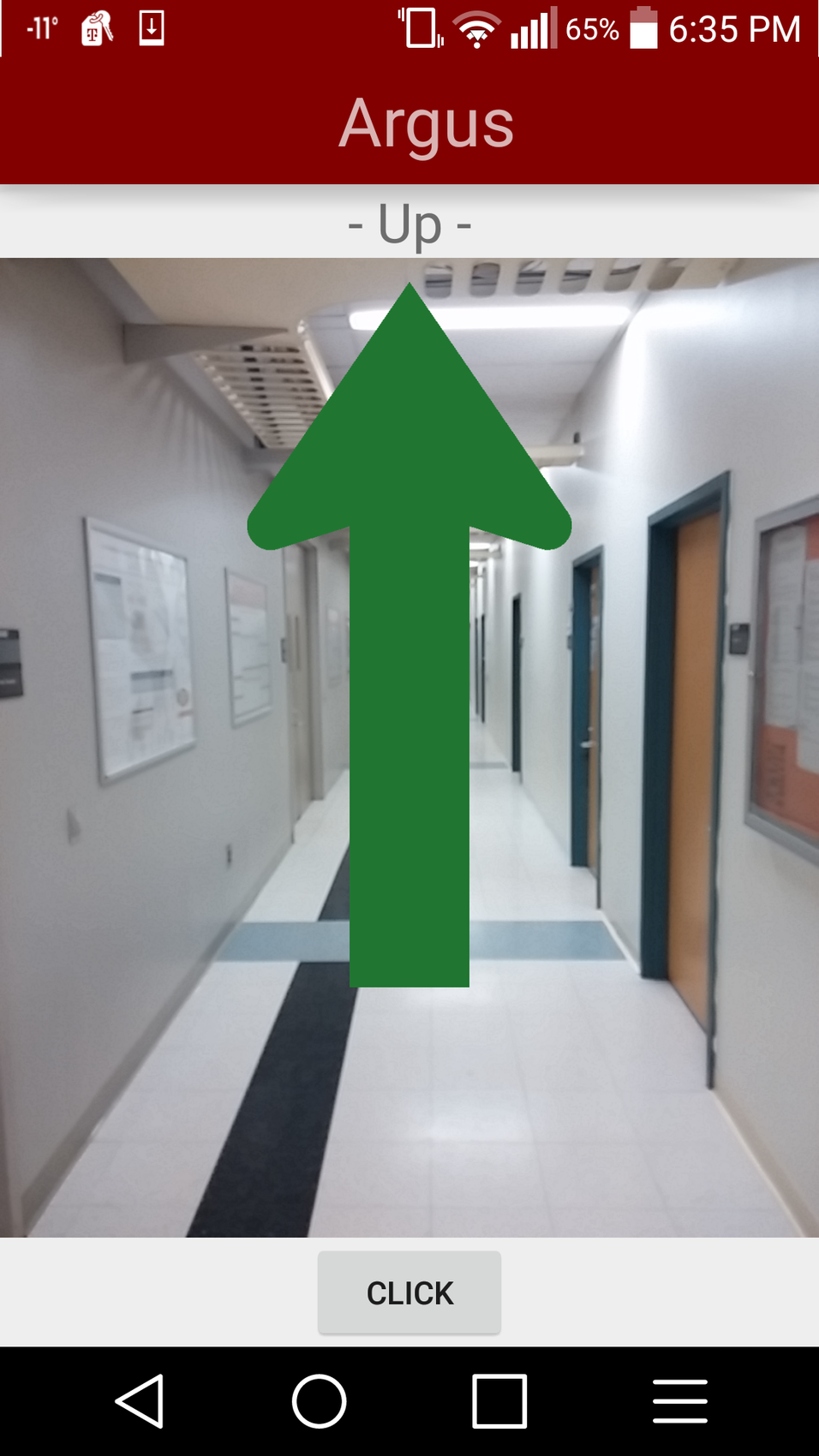}
            \caption{}
            \label{fig:arguscam}
        \end{subfigure}

        \caption{\label{fig:interface} Argus Android App: (a) Sensor options UI; (b) Initial instruction; (c) Camera UI and Feedback (Up/Down/Right/Left).}
        \vspace{-0.15in}
\end{figure}

\textbf{Smartphone Application: }Fig.~\ref{fig:interface} shows some screen-shots of our application. We enable the users to share other sensor information such as camera, microphone, GPS, accelerometer, gyroscope as this will help in the 3D reconstruction (Fig.~\ref{fig:argusconf}); Fig.~\ref{fig:arguspoint} shows the camera interface asking users to point towards incident zone. After capturing image, Fig.~\ref{fig:arguscam} shows the action to tilt UP (indicated by UP arrow) before taking the next picture. 
The app has an option to specify the upload server. In a real deployment, this will be the IP address of the central server in Fig.~\ref{fig:argus}.

\section{Performance Evaluation}\label{sec:evaluations}
\begin{figure*}[ht]
        \centering   
           \begin{subfigure}[b]{0.36\textwidth}
        		\centering
        		\includegraphics[width=1\textwidth]{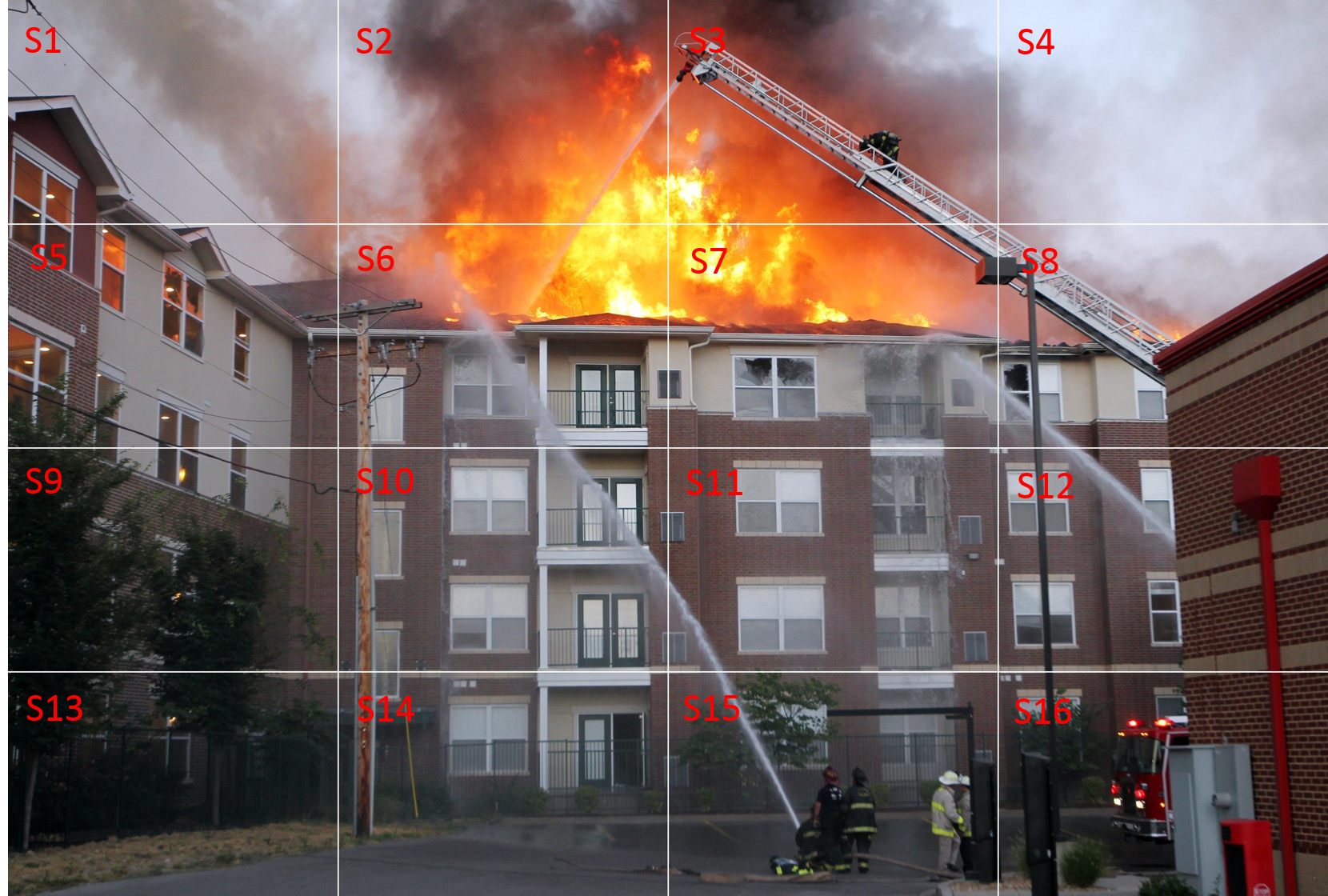}
        		\caption{}
        		\label{fig:buildingonfire}
        	\end{subfigure}
~
        \begin{subfigure}[b]{0.26\textwidth}  
            \centering 
            \includegraphics[width=1\textwidth]{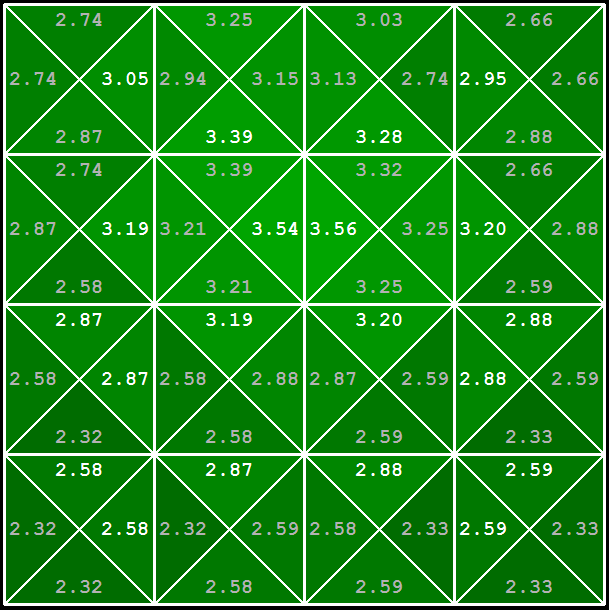}
            \caption{}
            \label{fig:qvalues}
        \end{subfigure}
~
        \begin{subfigure}[b]{0.26\textwidth}   
            \centering 
            \includegraphics[width=\textwidth]{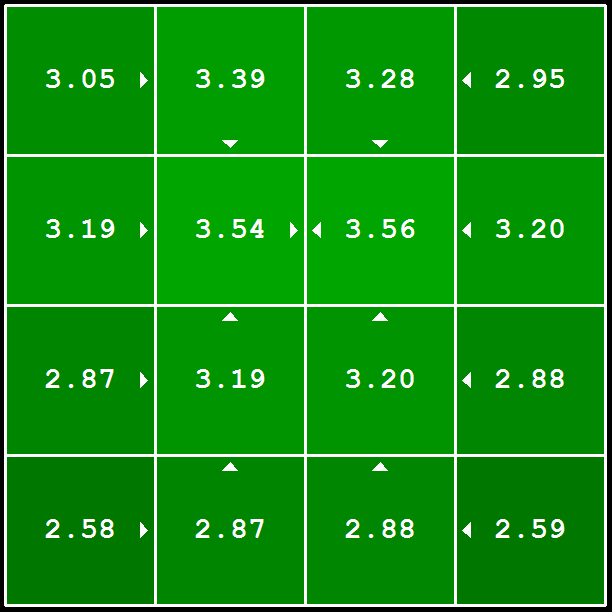}
            \caption{}
            \label{fig:values}
        \end{subfigure}
        \caption{\label{fig:distqlearn} (a) Incident scene with 4x4 MDP grid shown in white; (b) Final Q-Values calculated by our framework using distributed Q-learning; (c) Final values calculated by our framework with arrow heads showing the optimal actions for each state.}
        \vspace{-0.15in}
\end{figure*}

We evaluated our MARL framework based on distributed Q-learning in terms of key metrics such as percentage of time spent by agents on capturing the fire states (reward-states) vs. non-fire states as the parameters of the framework including the number of agents, decay rate of Boltzmann temperature, and learning rate are varied; also, we tested how well the algorithm adapts to random incidents. We coded the MARL framework in Python using distributed Q-learning on a 4x4 MDP grid shown in~Fig.~\ref{fig:buildingonfire} with $S13$ as the start state. 

\begin{figure*}[ht]
        \centering   
           \begin{subfigure}[b]{0.32\textwidth}
        		\centering
        		\includegraphics[width=1\textwidth]{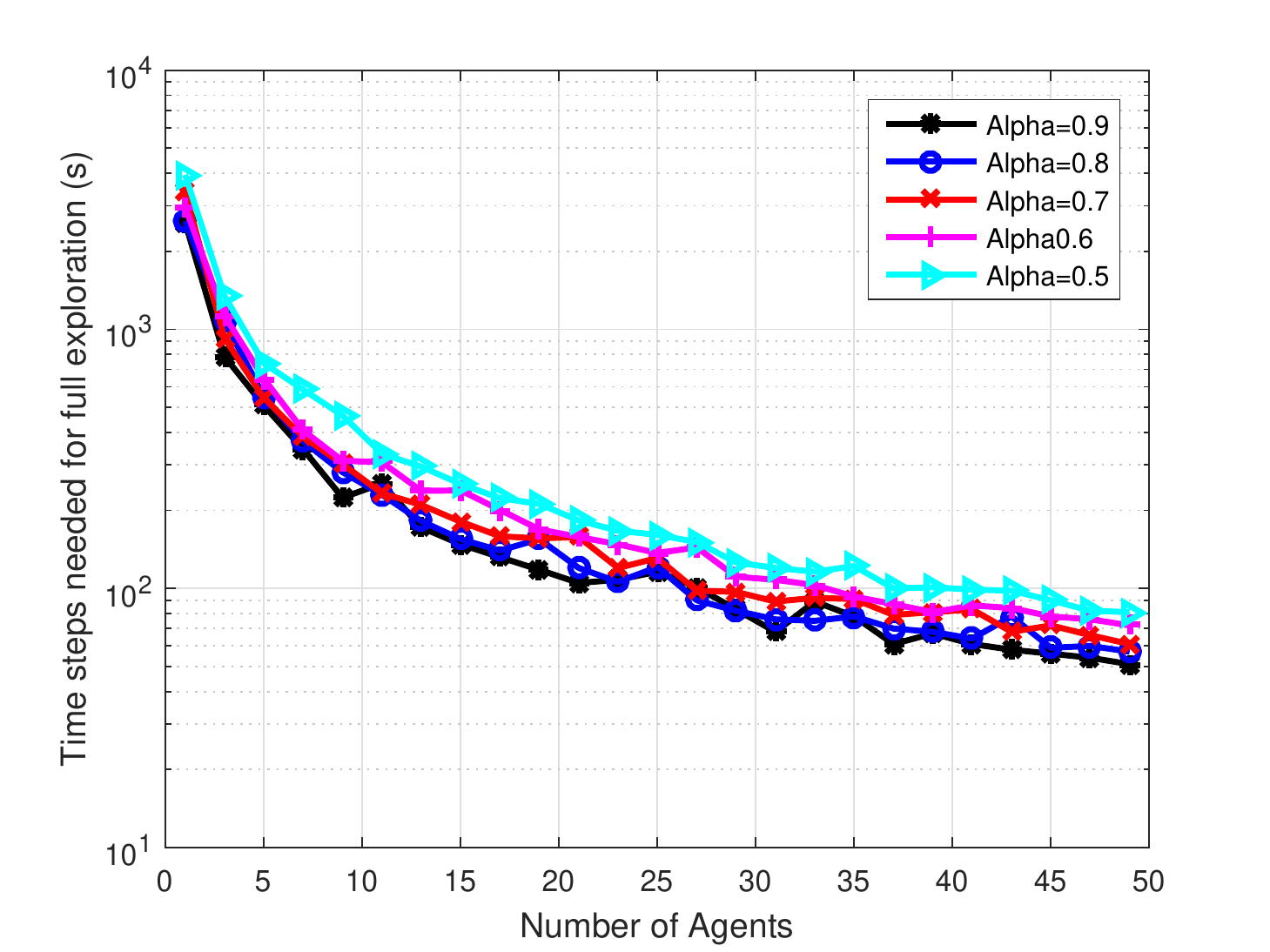}
        		\caption{}
        		\label{fig:expln-numagents}
        	\end{subfigure}
~
        \begin{subfigure}[b]{0.32\textwidth}  
            \centering 
            \includegraphics[width=1\textwidth]{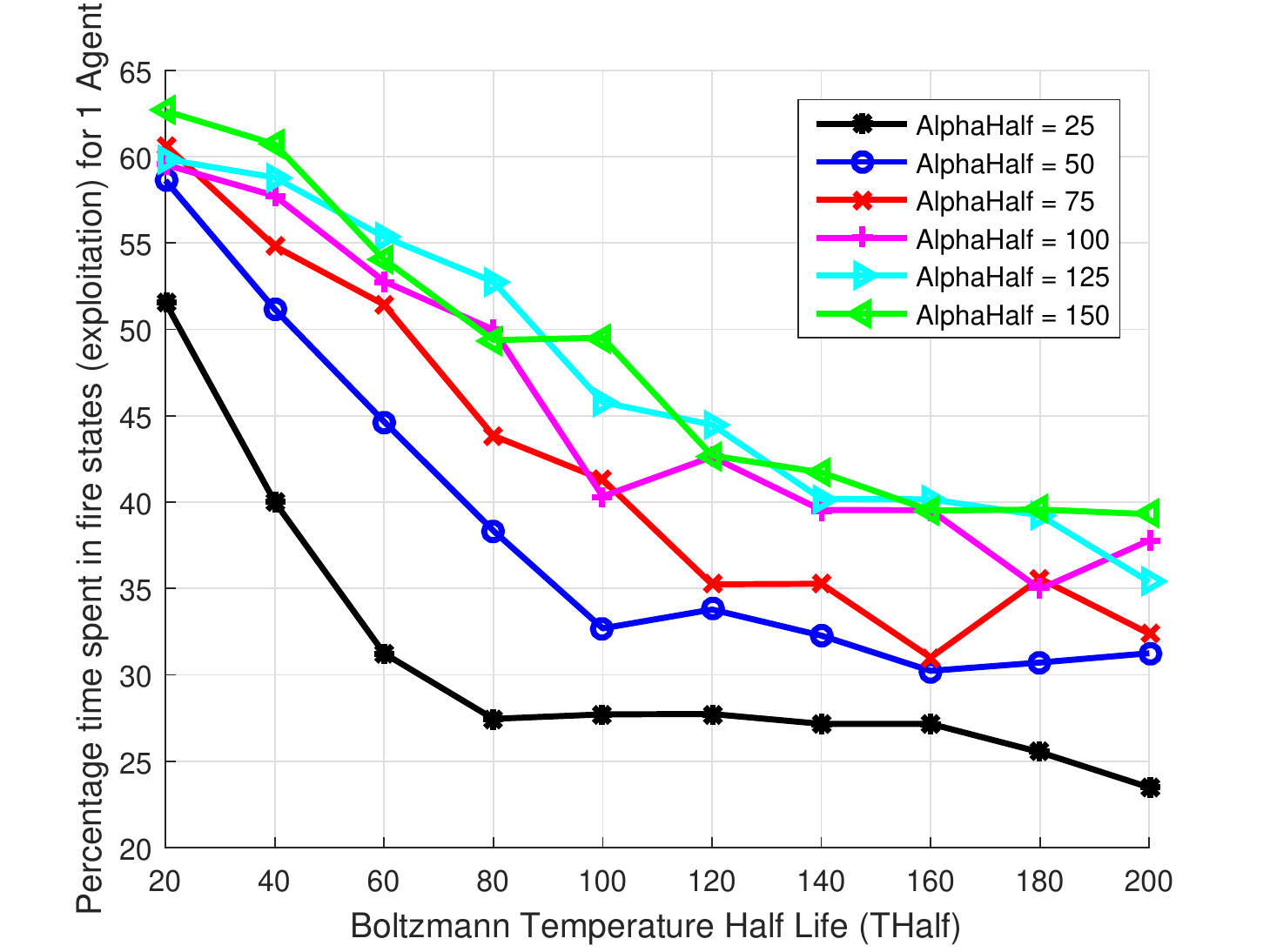}
            \caption{}
            \label{fig:exptn-vs-thalf-var-ahalf}
        \end{subfigure}
~
        \begin{subfigure}[b]{0.32\textwidth}   
            \centering 
            \includegraphics[width=\textwidth]{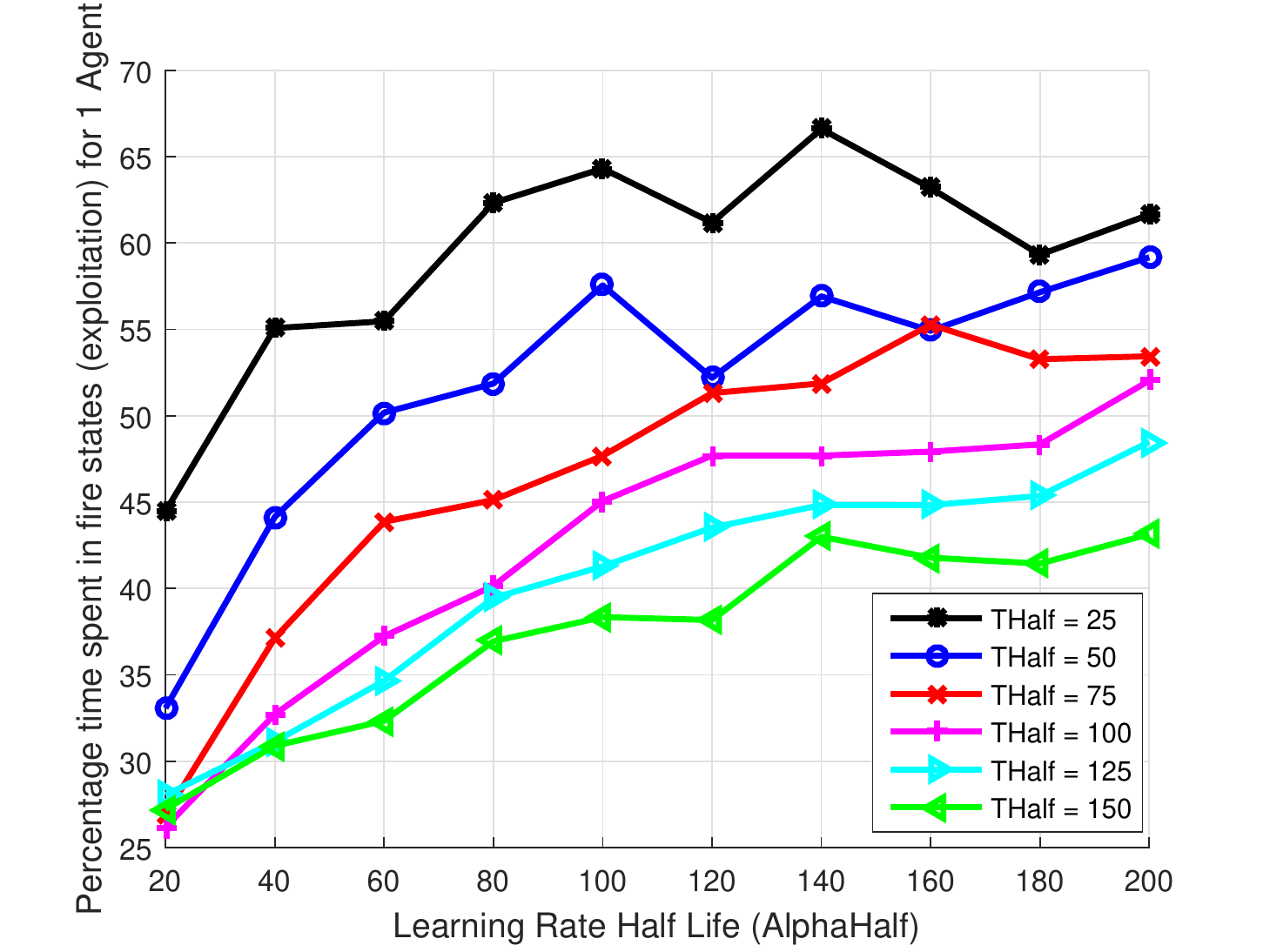}
            \caption{}
            \label{fig:exptn-vs-ahalf-var-thalf}
        \end{subfigure}
        \caption{\label{fig:agentsexplrn} (a) Time steps needed for full exploration as the number of agents increases for different learning rates; (b) Time spent in fire states [\%] vs. Temperature half life (1 Agent); (c) Time spent in fire states [\%] vs. Learning rate half life (1 Agent).}
        \vspace{-0.15in}
\end{figure*}
We remind the reader that the fire region will be divided into multiple MDPs where each MDP corresponds to a single surface of building, a corner of the building, or some other view of the incident scene. A rectangular grid will be overlaid on each of the views where the ensuing MDP will be solved. Fig.~\ref{fig:buildingonfire} shows an example of this scenario (the picture is sourced from the Internet). A 4x4 grid is overlaid on the incident zone, which means we have 16 states/cells in the MDP. Each agent's action in a state will include capturing a picture of the state (centering smartphone on the state) along with the physical action of panning/tilting. 
Since we do not have pictures of the incident scene in~Fig.~\ref{fig:buildingonfire}, we approximated the image of each state to the grid piece obtained by dividing the picture in~Fig.~\ref{fig:buildingonfire}, as per the white grid lines. Unlike in this simplified scenario, in reality we would have multiple pictures of each state captured by different agents at different times and the reward would be calculated as the percentage of fire pixels in those images. Using these approximated images (grid pieces), we calculated the percentage of fire pixels in them and then normalized it to obtain the rewards for these states. We can observe that only states S2, S3, S6, S7 have non-zero rewards. We ran the distributed Q-learning with discount factor of 0.9 until convergence of the values of the states. Note that all the agents share the same Q-values in our distributed Q-learning approach. The final Q-values and state values (which are maximum Q-values in each state) are shown in~Fig.~\ref{fig:qvalues} and Fig.~\ref{fig:values}, respectively. We can notice from~Fig.~\ref{fig:values} that fire states have higher values compared to the others and that the rewards propagate to surrounding states. Also, the optimal policy indicated by arrow heads leads the agent to the fire states where it stays most of the time.

\begin{figure*}[ht]
        \centering   
           \begin{subfigure}[b]{0.32\textwidth}
        		\centering
        		\includegraphics[width=1\textwidth]{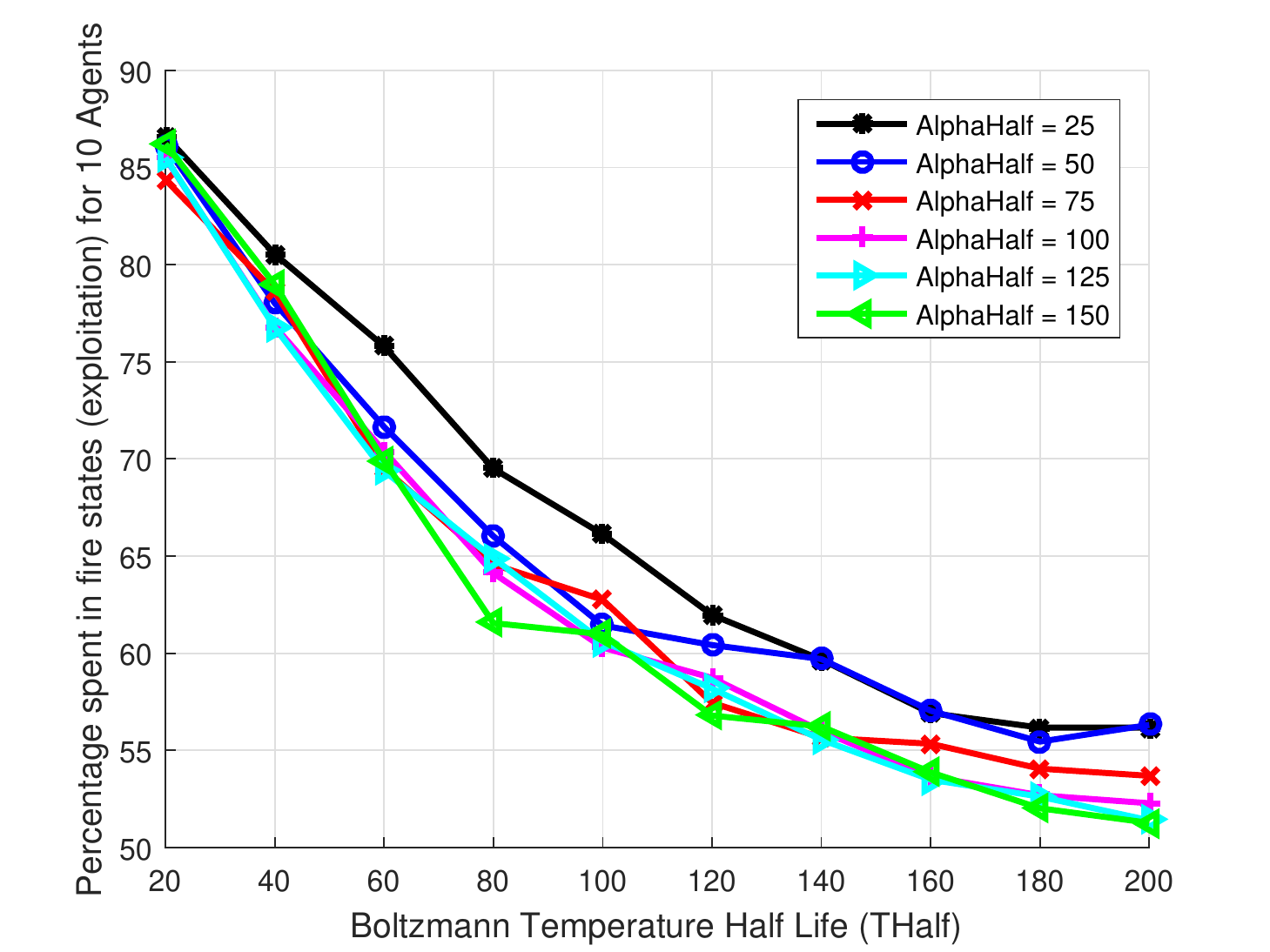}
        		\caption{}
        		\label{fig:exptn-vs-thalf-var-ahalf-10agents}
        	\end{subfigure}
~
        \begin{subfigure}[b]{0.32\textwidth}  
            \centering 
            \includegraphics[width=1\textwidth]{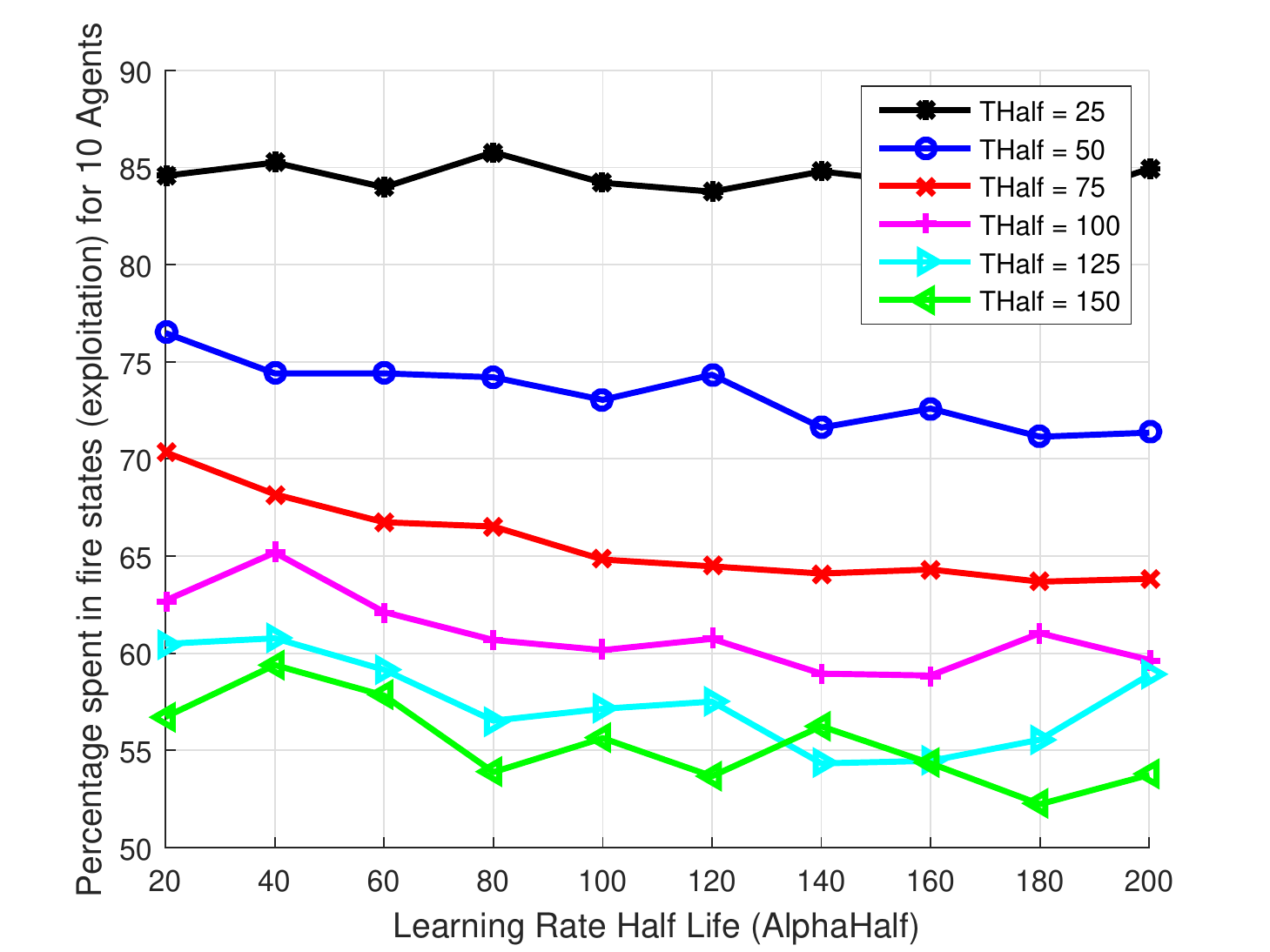}
            \caption{}
            \label{fig:exptn-vs-ahalf-var-thalf-10agents}
        \end{subfigure}
~
        \begin{subfigure}[b]{0.32\textwidth}   
            \centering 
            \includegraphics[width=\textwidth]{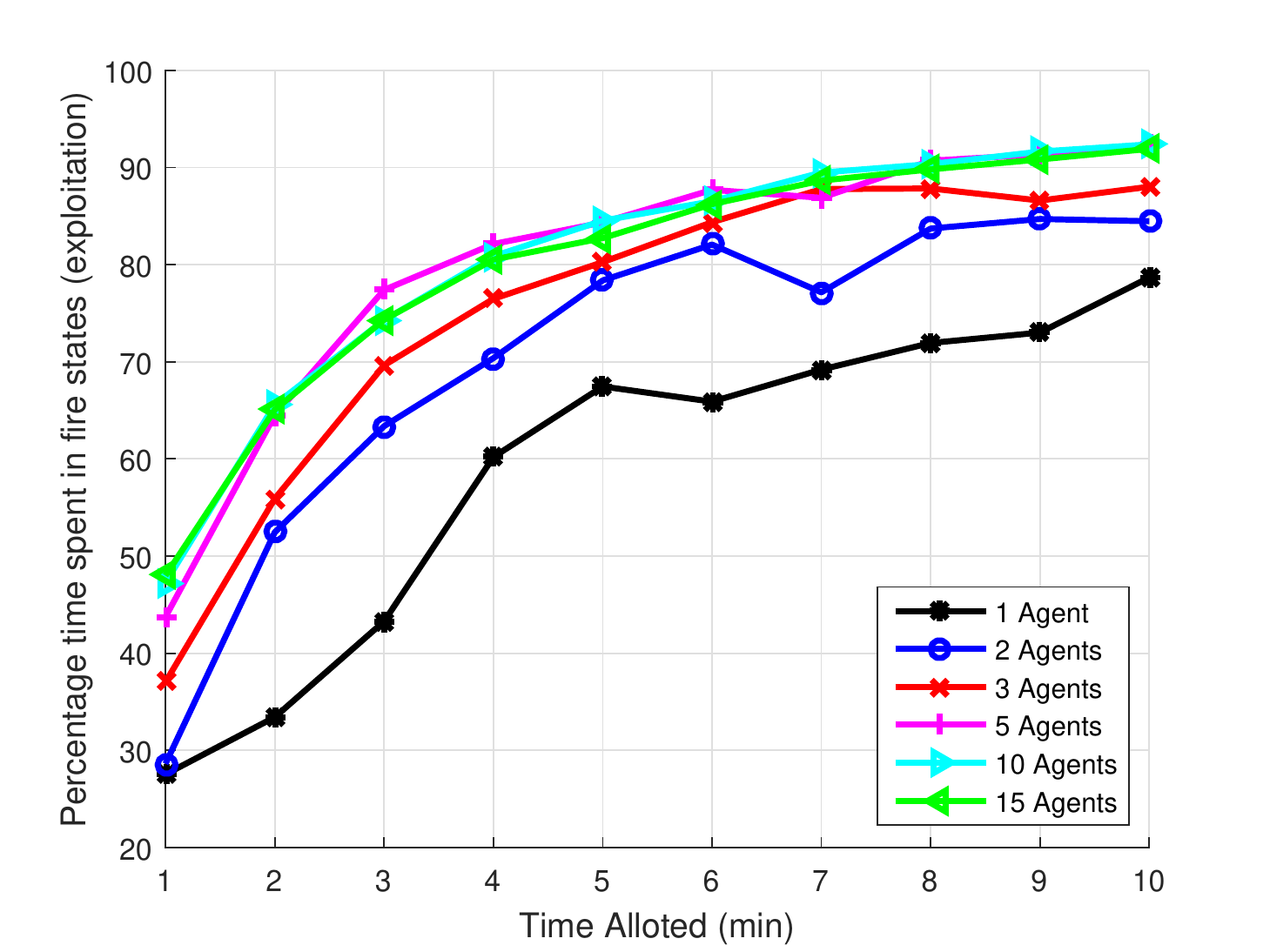}
            \caption{}
            \label{fig:exptn-vs-time}
        \end{subfigure}
        \caption{\label{fig:agentsexplrn10} (a) Time spent in fire states [\%] vs. Temperature half life (10 Agents); (b) Time spent in fire states [\%] vs. Learning rate half life (10 Agents); (c) Time spent in fire states [\%] as the time allotted to the agents increases ($T,\alpha = 50$).}
        \vspace{-0.15in}
\end{figure*}

\begin{figure}[ht]
        \centering   
           \begin{subfigure}[b]{0.15\textwidth}
        		\centering
        		\includegraphics[width=1\textwidth]{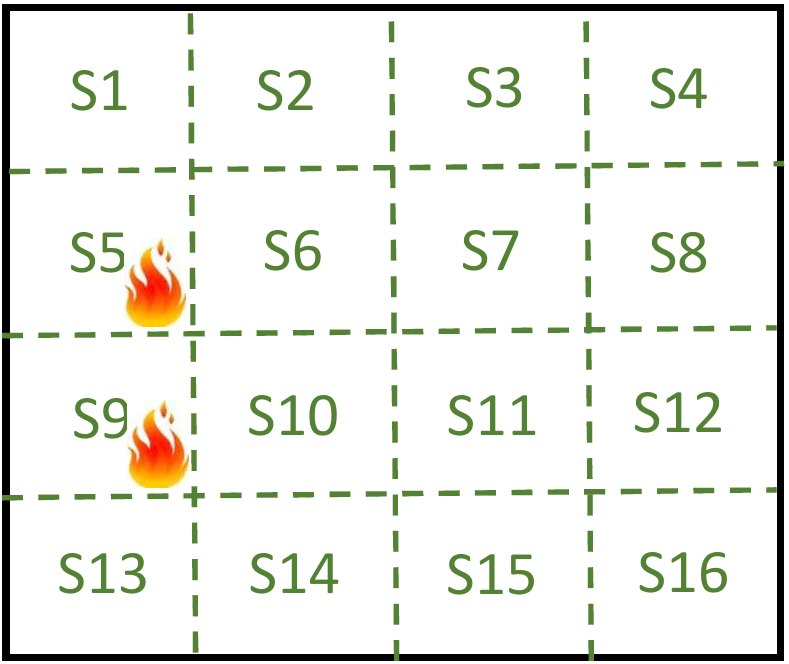}
        		\caption{}
        		\label{fig:s5s9}
        	\end{subfigure}
~
        \begin{subfigure}[b]{0.15\textwidth}  
            \centering 
            \includegraphics[width=1\textwidth]{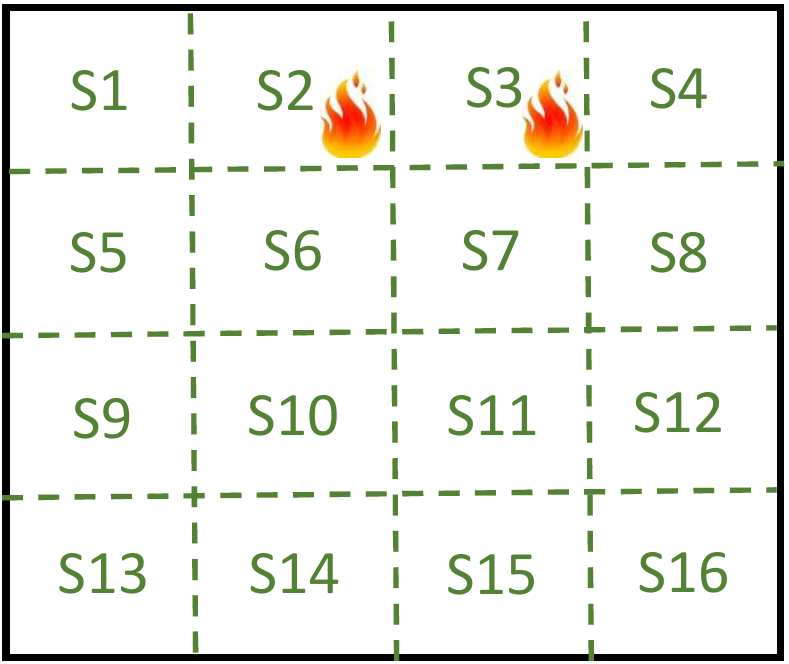}
            \caption{}
            \label{fig:s2s3}
        \end{subfigure}
~
        \begin{subfigure}[b]{0.15\textwidth}   
            \centering 
            \includegraphics[width=\textwidth]{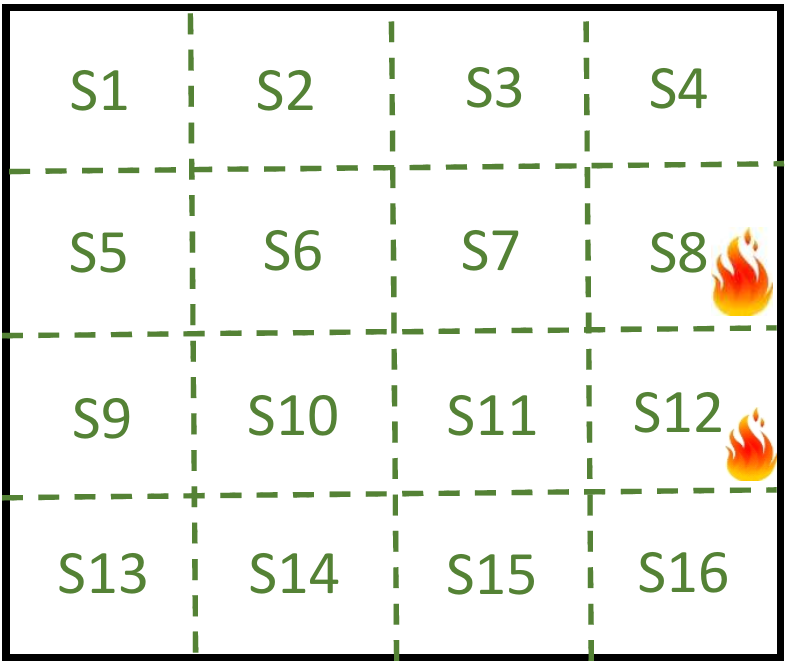}
            \caption{}
            \label{fig:s8s12}
        \end{subfigure}

           \begin{subfigure}[b]{0.15\textwidth}
        		\centering
        		\includegraphics[width=1\textwidth]{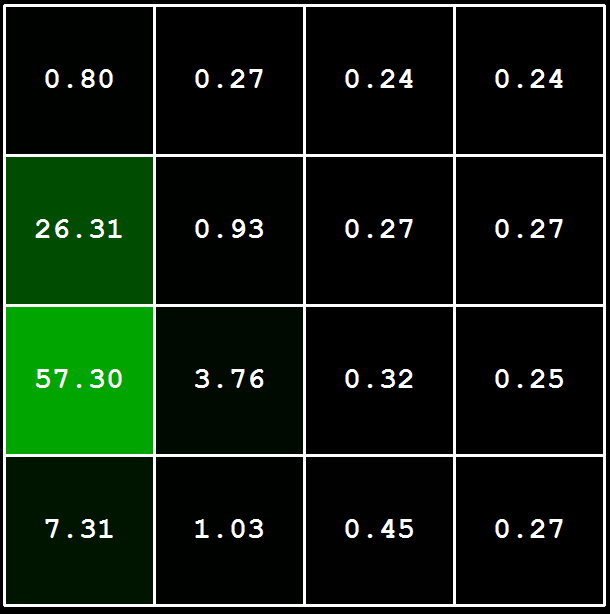}
        		\caption{}
        		\label{fig:design5x9}
        	\end{subfigure}
~
        \begin{subfigure}[b]{0.15\textwidth}  
            \centering 
            \includegraphics[width=1\textwidth]{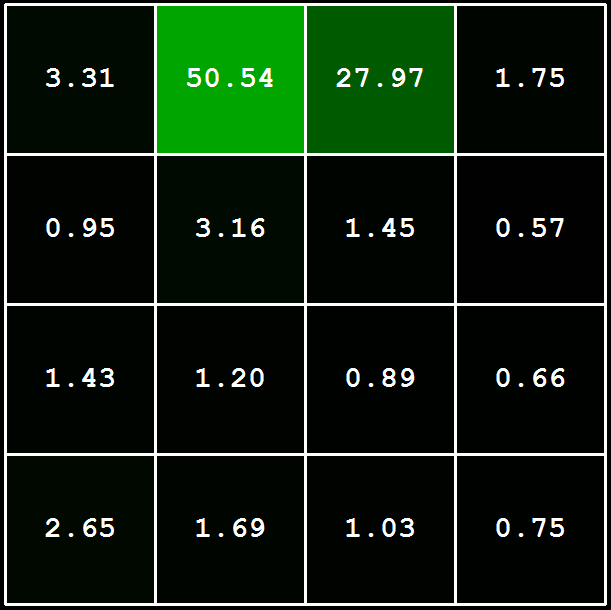}
            \caption{}
            \label{fig:design2x3}
        \end{subfigure}
~
        \begin{subfigure}[b]{0.15\textwidth}   
            \centering 
            \includegraphics[width=\textwidth]{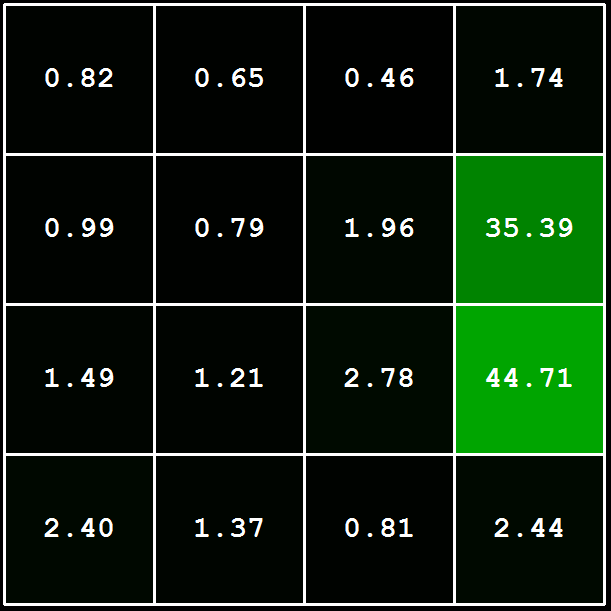}
            \caption{}
            \label{fig:design8x12}
        \end{subfigure}

        \caption{\label{fig:firepropagation} Random fire occurrences and model adaptation.}
        \vspace{-0.15in}
\end{figure}

\underline{Exploration vs. number of agents}: Fig.~\ref{fig:expln-numagents} shows the number of MDP time steps needed for full exploration of the 4x4 MDP grid shown in~Fig.~\ref{fig:buildingonfire} for different values of fixed learning rate, $\alpha$. We can see that the number of steps needed reduces drastically (from around 3000 to 80) as the number of agents increases, which shows the benefits of distributed Q learning and availability of multiple bystanders at the incident zone. We can also notice that $\alpha = 0.9$ is the best as the agents learn faster and take less number of steps to reach convergence. We also note that in practice there is no need to wait until full convergence of the values as policy converges faster than values. 
We also make an assumption that each time step in the Q-learning corresponds to about one second in wall-clock time. The decay in Boltzmann temperature $T$ (same for $\alpha$) is modeled as a radioactive decay, $T=T_{min} +(T_{max} - T_{min})e^{-\frac{\log 2}{T_{half}}t}$.

\underline{Time spent in fire states vs. $\alpha$ and $T$}: Fig.~\ref{fig:exptn-vs-thalf-var-ahalf} shows the percentage of time spent in fire states as $T_{half}$ is varied for different values of $\alpha_{half}$ for single-agent case with 180 time steps (3 minutes) (we have plotted the variation with different time steps in~Fig.~\ref{fig:exptn-vs-time}). We know that $T_{half}$ is inversely proportional to the decay rate. Large $T_{half}$ values correspond to more exploration and less exploitation, and viceversa. We can observe that the percentage of time spent on fire states reduces as $T_{half}$ increases. This is because there is more exploration as $T_{half}$ increases, thus reducing the time spent on fire states. We observe a similar behavior in~Fig.~\ref{fig:exptn-vs-ahalf-var-thalf}, which shows the percentage of time spent in fire states as $\alpha_{half}$ is varied for different values of $T_{half}$ also for a single agent with three minutes of allotted time. These results are averaged over 50 runs to account for randomicity present in the action selection that is based on the values of $T_{half}$. The ratio of exploitation to exploration can initially be around 75:25. Later this can be adaptively changed based on fire propagation. For example, if fire is spreading more rapidly, we would want to change this ratio to 90:10. However, we note that a single agent is not able to reach 70\% exploitation within the time allotted. Due to space constraints we have not included another figure, but the single agent is able to reach more than 75\% exploitation when the time allotted is increased to five minutes.
In order to show the benefits of multiple agents, we ran the same simulation with ten agents. The results are depicted in~Fig.~\ref{fig:exptn-vs-thalf-var-ahalf-10agents} and Fig.~\ref{fig:exptn-vs-ahalf-var-thalf-10agents}, which show that there is an improvement in the percentage of time spent in the fire states compared to the one-agent case. These results help us determine the values of the design parameters $T_{half}$, $\alpha_{half}$ based on the exploitation-exploration ratio preferences. For example, if we have ten agents and we prefer them to stay in fire regions for 75\% of time (in 3-minute window), we set $\alpha_{half}=T_{half}=50$.

\underline{Time spent in fire states vs. time allotted: }Fig.~\ref{fig:exptn-vs-time} shows the percentage of time spent in fire states as the time allotted to the agents increases from one to ten minutes for different number of agents. We have set $\alpha_{half}=T_{half} = 50$ based on our previous observation. We can note that the percentage increases both as the time allotted increases as well as the number of agents increases. However, we note a saturation behavior in both, implying that there is a minimum amount of time needed for exploration.

\underline{Model adaptation to random fire occurrences: }To evaluate our algorithm's ability to adapt to dynamic environments, we evaluated our MARL framework as the fire position changes every three minutes in a random manner, as shown in~Fig.~\ref{fig:firepropagation}. We can see that the system is able to adapt to change in fire positions by adaptively changing the percentage of time spent in fire states. We have enabled periodic exploration with resetting of parameters, i.e., Q-values to 0, $T_{half}=\alpha_{half}=50$.

\section{Conclusions and Future Work}\label{sec:discussions}
We presented Argus, a Multi-Agent Reinforcement Learning~(MARL) framework (implemented as a mobile application plus a backend server) for cooperative data collection to build a 3D mapping of the disaster scene using agents present around the incident zone to facilitate the rescue operations. As future work, we will incorporate local processing to handle network failures and develop privacy-preserving data-collection solutions; %
also, we plan to evaluate our framework in a pilot study.

\textbf{Acknowledgments}: We thank the Department of Homeland Security Science \& Technology Directorate (DHS S\&T) Cyber Security Division for their support %
under contract No.~D15PC00159.

\footnotesize
\bibliographystyle{ieeetr}
\bibliography{references}

\end{document}